\pdfoutput=1

\documentclass[sigconf]{acmart}

\usepackage{booktabs}   
\usepackage{subcaption} 

\usepackage[normalem]{ulem}

\usepackage{mathptmx} 

\usepackage[ruled,vlined,linesnumbered,resetcount]{algorithm2e}
\usepackage{amssymb}
\usepackage{amsmath}
\usepackage{physics}
\usepackage[flushleft]{threeparttable}

\newcommand{\cdl}{\textit{coupling degree list}}
\newcommand{\csm}{\textit{coupling strength matrix}}

\begin{document}
\title{Towards Efficient Superconducting Quantum Processor Architecture Design}

\author{Gushu Li}
\affiliation{
  \institution{University of California}
  \city{Santa Barbara}
  \country{USA}
}
\email{gushuli@ece.ucsb.edu}

\author{Yufei Ding}
\affiliation{
  \institution{University of California}
  \city{Santa Barbara}
  \country{USA}
}
\email{yufeiding@cs.ucsb.edu}

\author{Yuan Xie}
\affiliation{
  \institution{University of California}
  \city{Santa Barbara}
  \country{USA}
}
\email{yuanxie@ece.ucsb.edu}

\begin{abstract}

More computational resources (i.e., more physical qubits and qubit connections) on a superconducting quantum processor not only improve the performance but also result in more complex chip architecture with lower yield rate.
Optimizing both of them simultaneously is a difficult problem due to their intrinsic trade-off.
Inspired by the application-specific design principle, this paper proposes an automatic design flow to generate simplified superconducting quantum processor architecture with negligible performance loss for different quantum programs.
Our architecture-design-oriented profiling method identifies program components and patterns critical to both the performance and the yield rate.
A follow-up hardware design flow decomposes the complicated design procedure into three subroutines, each of which focuses on different hardware components and cooperates with corresponding profiling results and physical constraints. 
Experimental results show that our design methodology could outperform IBM's general-purpose design schemes with better Pareto-optimal results.

\end{abstract}

\maketitle

\thispagestyle{empty}

\section{Introduction}\label{sec:introduction}

As a promising computation paradigm, Quantum Computing~(QC) has been rapidly growing in the last two decades and found its strong potential in many important areas, including machine learning~\cite{harrow2009quantum, farhi2014quantum}, chemistry simulation~\cite{mcardle2018quantum, peruzzo2014variational}, etc.
In particular, the superconducting quantum circuit~\cite{devoret2013superconducting}
has become one of the most promising technique candidates for building QC systems~\cite{paik2011observation,barends2013coherent,chen2014qubit} due to the ever-increasing qubit coherence time, individual qubit addressability, fabrication technology scalability, etc. 
Towards efficient superconducting quantum circuit based QC system, significant research has recently been conducted, ranging from compiler optimization~\cite{shi2019optimized, murali2019noise} to periphery control hardware support~\cite{fu2017experimental, van2018electronic} and device innovation~\cite{koch2007charge, mckay2016universal}.   

\begin{figure*}[t]
\centering
\includegraphics[width=2.0\columnwidth]{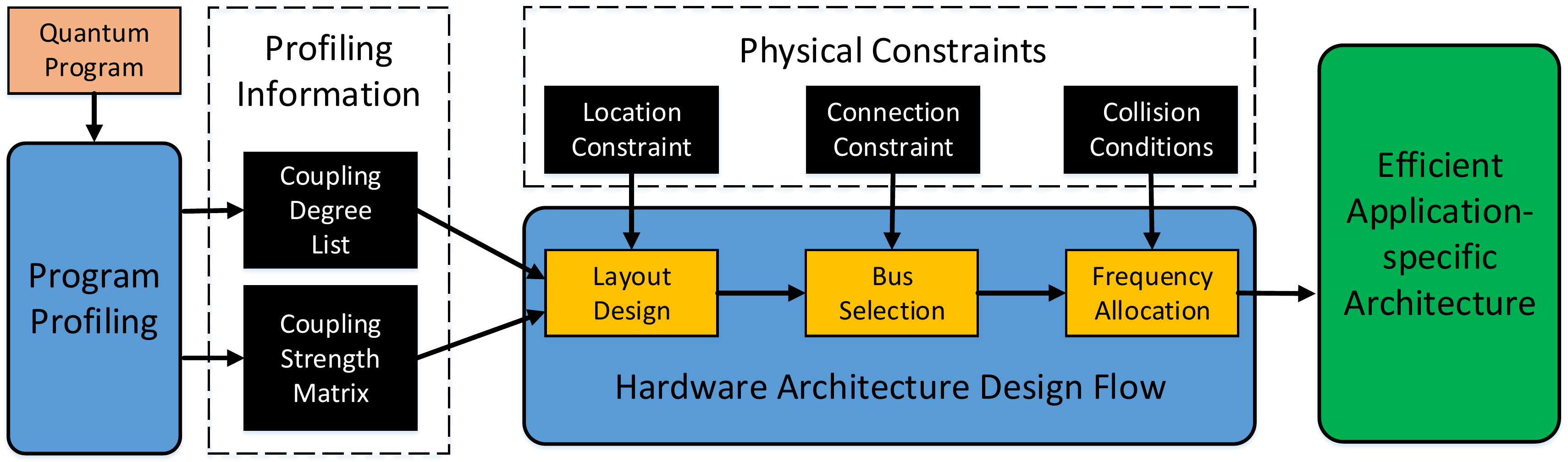}
\vspace{-5pt}
\captionsetup{justification=centering}
\caption{Overview of the Proposed Architecture Design Flow}
\vspace{-5pt}
\label{fig:overview}
\end{figure*}

Despite these system optimizations, the performance of a superconducting quantum processor is still highly limited by the amount of computation resource on it.
Researchers have been trying to integrate more qubits and qubit connections on one superconducting quantum processor substrate.
For example, IBM's first superconducting quantum chip on the cloud has 5 qubits with 6 qubit connections, while its latest published chip has 20 qubits with 37 qubit connections~\cite{cross2018validating}.
Increasing the number of physical qubits on a superconducting quantum processor allows programs with more logical qubits to be executed.  
Denser qubit connections can increase the overall chip performance by reducing the overhead of qubit mapping and routing ~\cite{siraichi2018qubit,zulehner2018efficient,li2019tackling,murali2019full}.



Nevertheless, more qubits and qubit connections will, unfortunately, increase the probability of defect occurrence on a chip, leading to lower yield rate and blocking future development of larger-scale superconducting quantum processor. 
For example, the yield rate of a 17-qubit chip can be lower than 1\% under IBM's state-of-the-art technology~\cite{rosenblatt2019enablement}.
Such a low yield rate comes from \textit{frequency collision}, a unique defect on superconducting quantum processors~\cite{magesan2018effective, brink2018device}. 
The frequencies of physically connected qubits may `collide' with each other when their values satisfy some specific conditions.
More qubit connections naturally increase the probability of frequency collision and lower the yield rate. 


To optimize both the yield rate and performance would be desirable, but it is difficult in general due to the inherent trade-off between these two objectives. 
Most previous efforts on them are direct device-level improvement~\cite{koch2007charge, mckay2016universal, kelly2015state, rosenblatt2019laser}, while little attention has been given to the architectural design of a superconducting quantum processor. 
This paper fills the gap by exploring the possibility of efficient \textit{application-specific architecture design} to reach an optimized balance between yield rate and performance.
We vision that an array of QC accelerators, each of which is tailored to a specific application, is much more likely to be adopted in the near term where computation resources are still limited before we can reach a universal quantum computer. 
Our design shares the same high-level spirit with the hardware architecture designs in classical computing (e.g., machine learning~\cite{chen2014diannao, han2016eie}, graph processing~\cite{ham2016graphicionado, ahn2016scalable}), but faces different scenarios because both the program patterns and the hardware design space are different in QC.

In particular, we highlight two key challenges to be addressed before the application-specific principle can be applied in superconducting quantum processor design. 
\textbf{First}, we need to 
identify and abstract the computation pattern of quantum programs that can guide the hardware architecture design. 
Prior quantum program analysis studies~\cite{javadiabhari2015scaffcc,ying2010quantum,ying2013verification,ying2013reachability,honda2015analysis,perdrix2008quantum} mainly focused on software or compiler optimization and cannot extract appropriate information for hardware architecture optimization.
\textbf{Second}, 
the abstracted computation pattern must give guidance to efficient architectural designs, which employ fewer computation resources with physical constraints satisfied to achieve both high yield rate and performance. 
Existing superconducting quantum processor design schemes cannot handle such irregular/complicated application-specific architecture design tasks~\cite{dallaire2016quantum,liebermann2017implementation,rosenblatt2019enablement, chamberland2019topological}.

To overcome these two challenges, we design a systematic design flow to automatically generate efficient superconducting quantum processor architecture designs for different quantum programs (shown in Figure~\ref{fig:overview}).  
We first identify two key computation patterns in quantum programs, \textbf{\cdl}~and \textbf{\csm}. A profiler is built to 
automatically extract them from an input quantum program. 
Both of them are critical to the program performance and hardware yield rate, and thus optimizing their underlying architecture support can potentially achieve a better balance between the performance and yield rate. 
We then propose an architecture design flow, which comes with three key subroutines, \textbf{\textit{layout design}}, \textbf{\textit{bus selection}}, and \textbf{\textit{frequency allocation}}. 
Each subroutine focuses on different hardware resources and must cooperate with corresponding profiling results and physical constraints.
We further propose an array of heuristics to ensure the scalability and effectiveness of the architecture search process. 
Empirical studies show that these heuristics can find `near-optimal' solution in the reduced search space.

In summary, this paper makes the following contributions:
\begin{itemize}
    \item We are the first to identify the optimization opportunity from the architecture level to push forward the balance between performance and hardware yield rate for superconducting QC processors. 
    \item We formalize an end-to-end design flow, equipped with a set of novel algorithmic primitives, to automatically generate a series of application-specific architectural designs under different hardware resource limits. 
    
    

    \item Comprehensive experiments show  that our design flow could outperform IBM's general-purpose designs with better Pareto-optimal results, e.g., magnitudes of yield improvement with negligible performance loss.
    
\end{itemize}




\section{Background}\label{sec:background}

In this section, we will introduce the necessary QC basics for understanding the following program profiling and superconducting quantum processor architecture design.

\subsection{QC Program Basics}
A quantum program can be represented in the well adopted quantum circuit model~\cite{nielsen2010quantum}.
We will start from the basic components in a quantum circuit and then illustrate how they compose a quantum circuit.

\textbf{Logical Qubit and Quantum Operation} 
A quantum program consists of some logical qubits as variables and some quantum operations which can modify the state of the qubits.
Qubit is the basic information processing unit in QC, which has two basis states denoted as $\ket{0}$ and $\ket{1}$.
One qubit can be not only the basis states themselves but also their linear combinations which can be depicted by a vector in the Hilbert space.
The state of the qubits can be modified by quantum operations.
The first type of quantum operation is unitary operation, also known as quantum gates in the circuit model, which can implement a unitary transformation on the qubit state.
Quantum gates can be applied on single qubit or multiple qubits.
The second type is measurement operation, which forces the qubits to collapse to basis states.

\textbf{Quantum Circuit} 
Quantum circuit is a model of QC in which the computation is a sequence of quantum gates and measurement operations.
The state of the qubits is first initialized and then manipulated by a sequence of operations.
Single-qubit gates and measurement operations are applied on individual qubits while two-qubit gates are applied on two logical qubits.
It has been proved that any multi-qubit gate can be decomposed into a series of single-qubit gates and CNOT gates (a specific two-qubit gate)~\cite{barenco1995elementary}. 
This is also the basic gate set directly supported on IBM's devices.
As a result, this paper assumes that the quantum circuit has been decomposed and gates with three or more qubits are not considered.


\subsection{Superconducting Quantum Circuit Basics}\label{sec:superq}
All the qubits and quantum operations in a quantum circuit must be implemented in a real physical QC system to execute the program.
In this paper, we focus on superconducting quantum processors with fixed-frequency Josephson-junction-based transmon qubits~\cite{koch2007charge} and all-microwave cross-resonance two-qubit gates~\cite{rigetti2010fully} adopted by IBM~\cite{rosenblatt2019enablement}.

\begin{figure}[t]
\centering
\includegraphics[width=0.9\columnwidth]{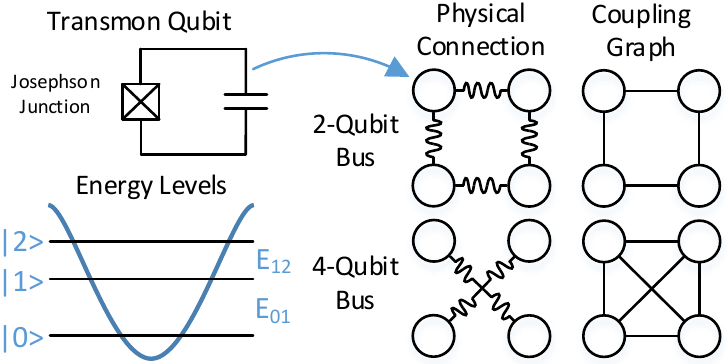}
\vspace{-5pt}
\caption{Superconducting Qubit and Connection}
\vspace{-5pt}
\label{fig:scqubit}
\end{figure}

\textbf{Physical Qubit and Frequency}
Figure~\ref{fig:scqubit} shows the physical circuit and energy levels of a transmon qubit~\cite{koch2007charge}. 
Due to the nonlinearity of the Josephson junction, the gaps between the energy levels in this quantum anharmonic oscillator are different, which allows us to use the ground state $\ket{0}$ and the first-excited state $\ket{1}$ as the computation basis without populating other states.
Suppose the energy gap between $\ket{0}$ and $\ket{1}$ for a qubit is $E_{01}$.
The \textit{frequency} of this qubit $f_{01}$ is defined as $f_{01} = E_{01}/h$, where $h$ is the Planck constant. 
Similarly, we use $f_{12}$ to represent the energy gap between $\ket{1}$ and $\ket{2}$.
For a typical qubit design with effective operations~\cite{magesan2018effective}, $f_{01}$ and $f_{12}$ are about $5GHz$ and $4.66GHz$, respectively. 
The anharmonicity of this qubit is defined to be $\delta = f_{12} - f_{01}$, which is $-340MHz$ under this typical design~\cite{chamberland2019topological,sheldon2016procedure}.

\textbf{Qubit Layout}
The superconducting physical qubits are confined on a 2-dimensional planar substrate.
Although the qubit placement can be flexible, major vendors fabricate the qubits in a regularized structure to ensure scalability and reduce the fabrication complexity.
For example, IBM's 16-qubit and 20-qubit chips \cite{IBMdevice} placed their qubits on the nodes of $2\times 8$ and $4\times 5$ lattices, respectively.
Google's 72-qubit chip placed its qubits on some nodes of an $11\times 12$ lattice~\cite{Google72Q}.

\textbf{Qubit Connection}
To enable two-qubit gates between two physical qubits, resonators, also known as qubit buses, are employed to connect nearby qubits~\cite{rigetti2010fully}. 
For examples, Figure~\ref{fig:scqubit} shows two types of commonly used buses.
The first one is a 2-qubit bus connecting two physical qubits.
The second one is a 4-qubit bus, which connects four physical qubits in a square together.
The coupling graphs of these two types of buses are shown on the right.
Compared with a 2-qubit bus, 4-qubit bus support two-qubit gates on not only the four qubit pairs on the edges but also two qubit pairs on the diagonals. 

\textbf{Qubit Mapping}
It is usually assumed that a two-qubit gate can be applied on arbitrary two logical qubits in a quantum program but some two-qubit gates may not be executable due to the limited qubit connection on a superconducting quantum processor.
On the hardware side, this problem can be relieved by employing more physical qubit connections so that two-qubit gates can be directly supported on more qubit pairs.
On the software side, a qubit-remapping compiler~\cite{maslov2008quantum} can resolve the dependency of the remaining unexecutable two-qubit gates while additional operations must be introduced with longer execution time and higher error rate.
Therefore, more physical qubit connections can help with the overall performance by allowing native two-qubit gates on more physical qubit pairs.



\textbf{Fabrication Variation}
Variation is inevitable when fabricating a superconducting quantum processor.
If a qubit is designed to have frequency $f$, the actual frequency after fabrication will be $f^\prime=f+n_f$, where $n_f$ satisfies Gaussian distribution $N(0,\sigma)$.
$\sigma$ is the fabrication precision parameter, which is around $130MHz \sim 150MHz$ under IBM's state-of-the-art technology~\cite{rosenblatt2019enablement}.
Such noise makes it hard to predict the post-fabrication frequency precisely, which brings the probability of frequency collision. 

\begin{figure}[h]
\centering
\vspace{-5pt}
\includegraphics[width=1.0\columnwidth]{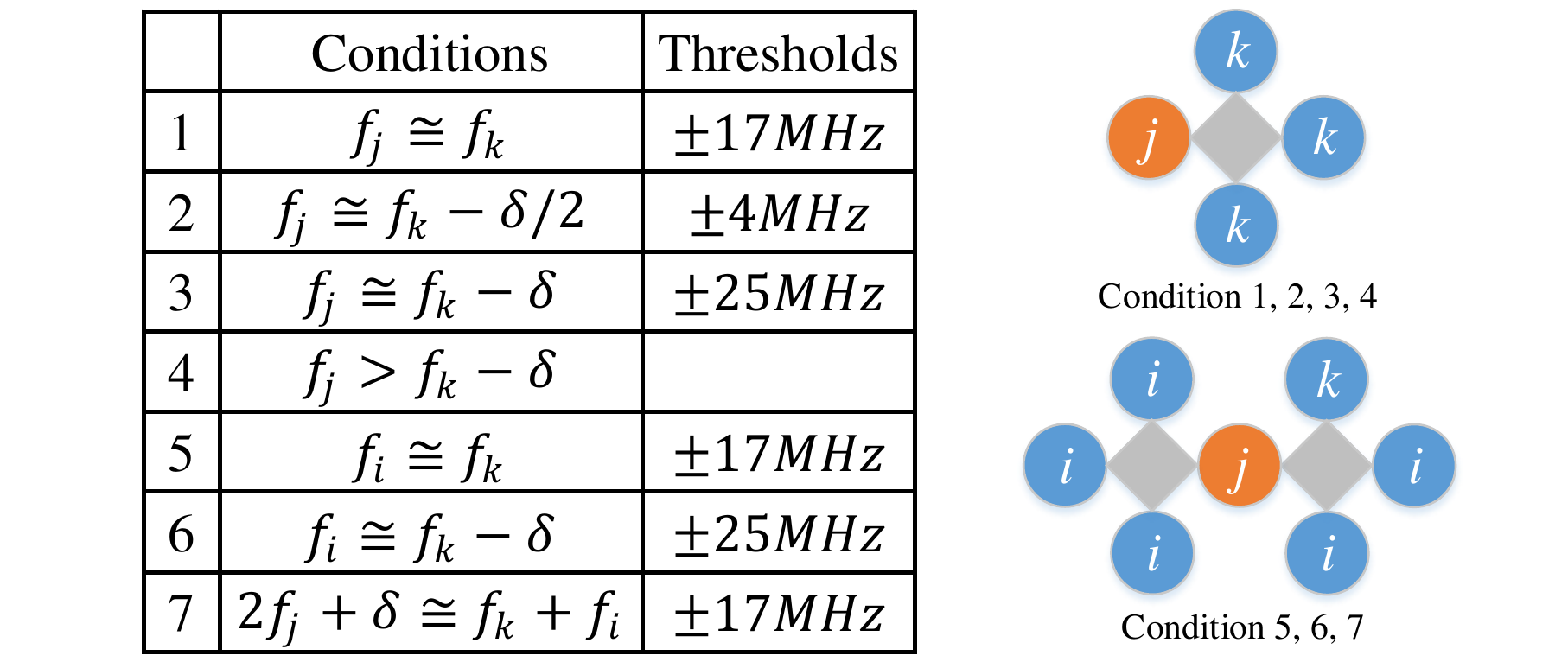}
\vspace{-15pt}
\caption{Frequency Collisions Conditions~\cite{rosenblatt2019enablement,brink2018device}}
\vspace{-5pt}
\label{fig:collisions}
\end{figure}

\textbf{Frequency Collision}
When two or three qubits are connected, \textit{frequency collision} may happen and cause defects on the device.
Figure~\ref{fig:collisions} summaries seven qubit frequency collision conditions in IBM's devices~\cite{rosenblatt2019enablement, brink2018device}. 
On the left is a table showing the conditions and thresholds of different collision situations. 
Condition 1, 2, 3, and 4 involve two connected qubits (j and k).
Condition 5, 6, and 7 involve three qubits of which two qubits (k and i) both connect to the other qubit j.
The approximate equations and the corresponding thresholds determine whether one frequency collision happens.
For example, if qubit $j$ and $k$ are connected and $|f_j - f_k| < 17MHz$, then the first condition is satisfied and frequency collision occur.
Note that the fourth condition has no threshold because it is an inequality rather than an approximate equation.
On the right is a graphical illustration, showing the geometric locations of the qubits that may have frequency collisions of different conditions in two subfigures.
Each circle represents one qubit and the gray square represent a 4-qubit bus connecting the four surrounding qubits.

\begin{figure*}[t]
\centering
\includegraphics[width=2.0\columnwidth]{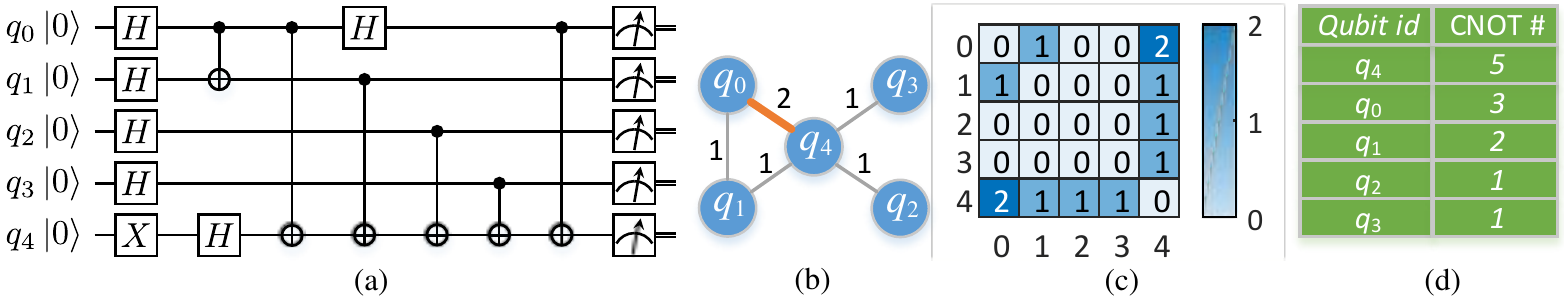}
\vspace{-5pt}
\captionsetup{justification=centering}
\caption{Example of the Profiling Method}
\vspace{-5pt}
\label{fig:profiling}
\end{figure*}

\section{Quantum Program Profiling}\label{sec:profiling}

The first step towards the development of an architecture-specific quantum processor for both high performance and yield rate is to determine what program information we should focus on. 
There are several different types of components in a quantum circuit but not all of them will significantly affect the hardware design.
Our target program component(s) should satisfy two conditions. 1) The component's execution is a performance bottleneck which can be dramatically improved with optimized hardware support.
2) The component's required hardware should significantly affect the yield rate.

We found that two-qubit gates can be a key factor to bridge performance and yield.
To execute two-qubit gates on a quantum processor with limited qubit-to-qubit coupling, a large number of additional operations are introduced to satisfy their dependencies.
But implementing two-qubit gates on two physical qubits require on-chip qubit connections which can lower the yield rate through increasing the probability of frequency collision. 
Therefore, we give logical qubits and qubit pairs priorities based on the number of involving two-qubit gates to help with the following architecture design.
Critical qubits and qubit pairs will have more hardware support to improve the efficiency of the generated architectures.





These remaining components, single-qubit gates, initialization, and measurement operations, do not involve qubit-to-qubit interactions and all happen locally on individual qubits when they are implemented on hardware. 
As a result, hardware support for these components will not affect the chip yield through frequency collision. 

\subsection{Profiling Method}
As discussed above, our profiling will focus on the logical qubits and the two-qubit gates.
Figure~\ref{fig:profiling} shows an example to illustrate the profiling procedure.
Suppose we have a quantum circuit as shown in Figure~\ref{fig:profiling} (a).
It has 5 logical qubits denoted by $q_{0,1,2,3,4}$.
All of them are initialized to be $\ket{0}$.
Then some single-qubit gates and two-qubit gates are applied.
Measurement operations are at the end.

We first ignore all single-qubit gates, initialization, and measurement operations.
Then we create a logical coupling graph, in which each vertex represents one logical qubit in the circuit.
Two vertices are connected by an undirected edge if there exists two-qubit gates applied on the two corresponding logical qubits.
The weight of an edge is the number of two-qubit gate instances on the two connected vertices.
In this example, Figure~\ref{fig:profiling} (b) shows the generated graph for the example circuit.
The weight of the edge between vertex $q_{0}$ and vertex $q_{4}$ is 2 since there are two two-qubit gates on $q_0$ and $q_4$. 
For all other edges, the weight is 1 because there is only one two-qubit gate on each of those qubit pairs.
The first profiling result is the weighted adjacency matrix of the logical coupling graph, namely the \csm.
The element with indices $(i,j)$ represents the number of two-qubit gates between $q_i$ and $q_j$.
Figure~\ref{fig:profiling} (c) shows the \csm~for the example circuit. 
Note that \csm~is always a symmetric matrix.

The second result is \cdl. 
For each qubit, we sum the weights of edges that connect to its corresponding vertex and define the number of two-qubit gates applied on it as the \textit{coupling degree} of one qubit.
If one qubit is associated with more two-qubit gates in a quantum circuit than other qubits, this qubit will use the physical qubit connections more frequently when executing on the chip.
Naturally, we should pay more attention to those qubits with larger coupling degree.
Therefore, all qubits are placed in a sorted list, namely the \cdl.
Figure~\ref{fig:profiling} (d) is the \cdl~in this example.
The first one in this list is $q_4$ because it has the largest coupling degree.
All qubits are in a descending order.


\begin{figure}[h]
\centering
\vspace{-3pt}
\includegraphics[width=1.0\columnwidth]{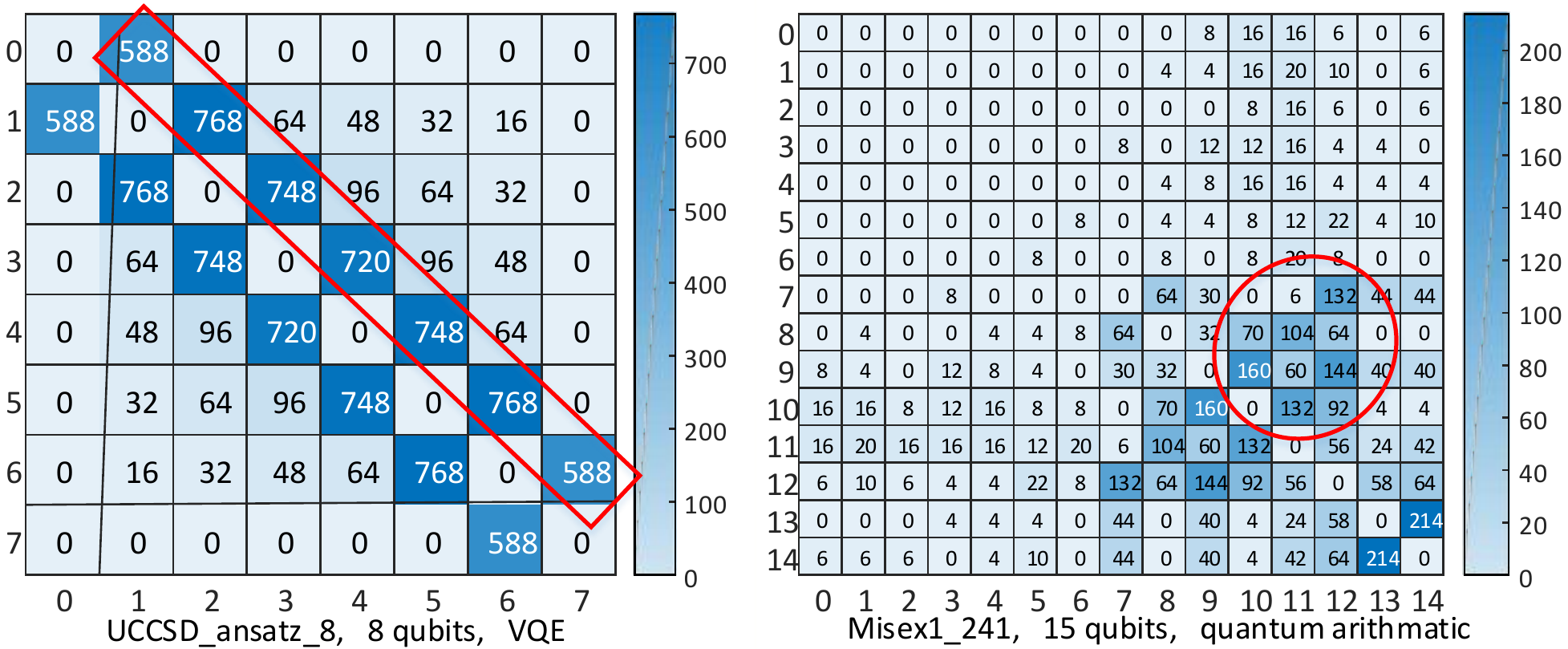}
\vspace{-17pt}
\captionsetup{justification=centering}
\caption{Qubit Coupling Strength Patterns for Two Programs}
\label{fig:motivatingexample}
\end{figure}
\subsection{Gate Pattern Examples}



In this section, we show the existence of distinct two-qubit gate patterns and discuss the opportunity for application-specific architecture design with two examples.
Figure~\ref{fig:motivatingexample} shows their \textit{coupling strength matrices}.
On the left is an 8-qubit UCCSD ansatz for VQE, a quantum simulation algorithm~\cite{peruzzo2014variational}.
The high coupling strength qubit pairs form a chain structure marked by a red rectangle.
$Q_0$ and $Q_1$ have a large number of two-qubit gates between them, as well as $\{Q_{1}Q_{2}$, $Q_{2}Q_{3}$, $\cdots$, $Q_{6}Q_{7}\}$. 
For other qubit pairs, the coupling strength is much lower~(only about 10\%).
On the right is a 15-qubit quantum arithmetic function~\cite{wille2008revlib}.
The coupling strength among $Q_0Q_1\cdots Q_{5}$ are $0$ since there are no two-qubit gates on any two of them.
However, there is a large number of two-qubit gates where one qubit is in the set $Q_{7,8,9,10}$ and the other qubit is in the set $Q_{10,11,12}$~(marked by a red circle).
The analysis of these two motivating examples provides us two observations:
\begin{enumerate}
    \item The numbers of two-qubit gates on different logical qubit pairs can vary dramatically in a real quantum program. 
    \item Different types of quantum programs can have different two-qubit gate patterns.
\end{enumerate}
These observations suggest that quantum processors can be customized for different programs with different patterns. 
An efficient architecture can focus on supporting the high-density coupling in a quantum program to reduce the number of connections on-chip.
For example, a quantum processor with an 8-qubit chain structure (8 qubits and 7 qubit connections) can immediately support most of the two-qubit gates in the 8-qubit UCCSD ansatz program.
The rest two-qubit gates can be supported through remapping without introducing too many additional operations because the total number of the remaining two-qubit gates is relatively small.
Such application-specific QC accelerators with simplified architectures can be a more realistic goal in the near term than a general-purpose quantum processor with a large number of hardware resources.



\section{Architecture Design}\label{sec:organization}

After a quantum circuit is profiled, a straightforward quantum processor architecture for such a circuit is to organize the on-chip qubits and qubit connections directly based on the logical coupling graph. 
However, we must consider the physical constraints for a practical architecture. 
For example, a logical coupling graph may not be perfectly fabricated on hardware since the allowed connections among superconducting qubits are very limited.
Moreover, we hope to improve the yield rate by delivering architecture designs with fewer hardware resources. 
Therefore, the proposed hardware design flow must not only invest more hardware resource on frequent operations based on the profiling results, but must also obey the physical constraints on the hardware components arrangement.

To accomplish such a complicated task in a scalable way, we decouple the hardware design procedure into three subroutines and each subroutine focuses on different architecture components, i.e., qubit layout, connection, and frequency.
For each subroutine, we first review the difficulty and the physical constraints considered. 
Then we discuss the design objectives, and how they are achieved in the proposed design algorithms.




\subsection{Layout Design}

\begin{figure}[b]
\centering
\includegraphics[width=1.0\columnwidth]{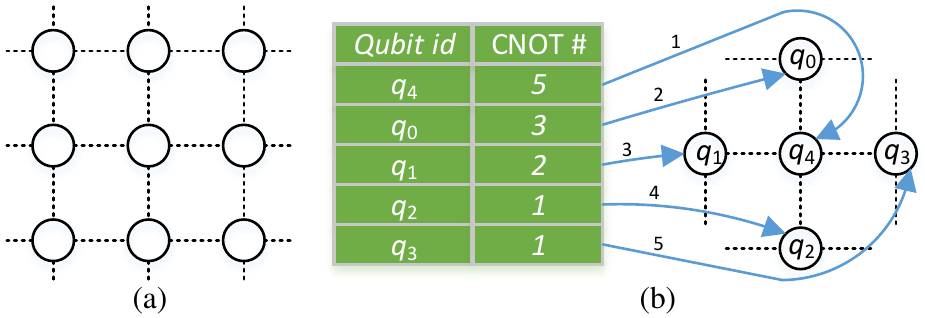}
\vspace{-15pt}
\captionsetup{justification=centering}
\caption{(a) Empty Lattice (b) Qubit Placement Example}
\vspace{-5pt}
\label{fig:layoutdesign}
\end{figure}

The first step is to determine where to place the qubits.
To ensure scalability and modularity, we follow the convention from major vendors introduced in Section~\ref{sec:background} and will only place qubits on the nodes of a 2D lattice.  
We start from a large 2D lattice, in which each node is initialized to be empty (Figure~\ref{fig:layoutdesign} (a)).
Then physical qubits can be placed in the empty nodes and one node can contain at most one qubit.

There are many ways to place a given number of qubits on a 2D lattice.
For example, 16 qubits can constitute a $4\times4$ lattice, a $2\times8$ lattice, or other more irregular structures.
But we need to select one qubit layout that is most suitable for executing the program, i.e., most operations can be directly supported or indirectly supported with low overhead.
The objectives of this qubit layout design subroutine are summarized as follows. 
\begin{itemize}
    \item Since we need to consider the profiling information, we create a pseudo mapping between logical qubits in the profiled program and the physical qubits in hardware architecture to be delivered. For two logical qubits with a large number of two-qubit gates between them, we hope to place their corresponding physical qubits in adjacent nodes so that later those two-qubit gates can be directly supported by the connection between the two physical qubits.
    \item One physical qubit can only have a limited number of directly connected qubits. For those two-qubit gates that cannot be directly supported, we hope to reduce the amount of additional operations introduce for remapping the qubits.
\end{itemize}



\begin{algorithm}[t]
\SetAlgoLined
 
\KwIn{\cdl~ $L$, \csm~ $M$}
\KwOut{Geometric coordinates of placed qubits}
Place the qubit with the largest coupling degree in $L$ at one node with coordinate $(0,0)$\;
 $R$ = all the qubits remaining; \tcp{R is the set of qubits that has not been placed yet.}
 \While{$R$ is not empty}{
 
 \tcc{Find the next qubit to place}
 
 $qubit\_candidate\_list = \varnothing$ \;
 \For{\textbf{q} in R}{
    \If{\textbf{q} is connected to any placed qubits}{
        $qubit\_candidate\_list.append(\textbf{\textit{q}})$\;
    }
 }
 Find the qubit \textbf{\textit{q}} with the largest coupling degree in $qubit\_candidate\_list$\;
 
 $node\_cost = [~]$\;
 \tcc{Determine the placement location}
 \For{\textbf{location} of the nodes that are empty and connected to at least one occupied node}{
 \tcc{Heuristic Cost function}
 $node\_cost[\textbf{\textit{location}}] = \sum\limits_{q^\prime \in \textbf{\textit{q}}.neighbors}{M[\textbf{\textit{q}},q^\prime]
 *distance[\textbf{\textit{location}},q^\prime.node]}$

 }
   \tcc{$q^\prime$ must be placed neighbor qubits}
 Place \textbf{\textit{q}} in the \textbf{\textit{location}} with the minimal score\;
 $R.remove(\textbf{\textit{q}})$\;
 
}

\caption{Qubit Placement on 2D Lattice}
\label{alg:layout}
\end{algorithm}

We propose a \textit{coupling-based} qubit placement algorithm to determine the geometric locations of the qubits on a 2D lattice (pseudocode shown in Algorithm~\ref{alg:layout}).
We illustrate the algorithm with an example in Figure~\ref{fig:layoutdesign}.
First, we put the first qubit in the \cdl, $q_4$, on one node of the 2D lattice.
Since the initial 2D lattice is empty, the location of $q_4$ does not matter.
We set the geometric coordinate of the first qubit to be $(0,0)$ and then place the rest qubits around $q_4$.
$q_4$ has four neighbors, $q_{\{0,1,2,3\}}$, in the logical coupling graph. 
We need to select the next one to place.
By checking the \cdl, we can see that $q_0$ is the one with the largest coupling degree.
The node occupied by $q_4$ has four equivalent adjacent nodes and we can place $q_0$ on any of them. 
In this example, we select the node on the north of $q_4$ with coordinate $(0,1)$.
Such an algorithm design ensures that the strongly coupled qubit pairs are given higher priority and placed on adjacent nodes, accomplishing the first objective mentioned above.

Then we need to place $q_1$ since its coupling degree is larger than that of $q_2$ and $q_3$.
$q_1$ is connected to both $q_4$ and $q_0$ so that we need a more sophisticated way to evaluate all potential nodes for $q_1$.
We use the function in line 13 of Algorithm~\ref{alg:layout} to find the node that can make $q_1$ close to its strong coupled neighbors in the logical coupling graph.
This function is the summation over all $q_1$'s placed neighbors. 
Each term in the summation is the product of the coupling strength between $q_1$ and one logical coupling neighbor $q^\prime$ and the Manhattan distance between the evaluated node location and the location of $q^\prime$.
After evaluating all the empty nodes that are adjacent to placed nodes $q_4$ and $q_0$, we will find that the nodes on the east and west of $q_4$ are the best ones because they are closest to $q_4$ but not far away from $q_0$.
Here we select the one on the west of $q_4$ with coordinate $(-1,0)$.
This summation function can help reduce the number of operations for later remapping and achieve the second design objective.

The remaining qubits can be placed in a similar procedure until all the qubits have been placed on the 2D lattice.
In this example, $q_2$ and $q_3$ are placed on the nodes with coordinates $(0,-1)$ and $(1,0)$, respectively.
All the qubits have their locations~(coordinates) on a 2D lattice where we can fabricate one physical qubit on each occupied node.
Finally, the nodes with no qubits are removed.

\subsection{Bus Selection}

In the second step, we need to connect the placed physical qubits to enable two-qubit gates.
The difficulty comes from the large size of the design space.
For $N$ qubits, there are $N(N-1)/2$ distinct qubit pairs.
Any of them can be either connected or disconnected so that there are $2^{N(N-1)/2}$ different cases.
Even after considering the nearest-neighbor coupling constraint in which one qubit can only connect with few qubits around it on the lattice, the size of the design space is still $O(exp(N))$.
More importantly, more qubit connections will improve the performance but lower the yield rate in general so that we need to identify those connections with the most potential performance benefit in a very large design space.

\begin{figure}[b]
\centering
\includegraphics[width=.8\columnwidth]{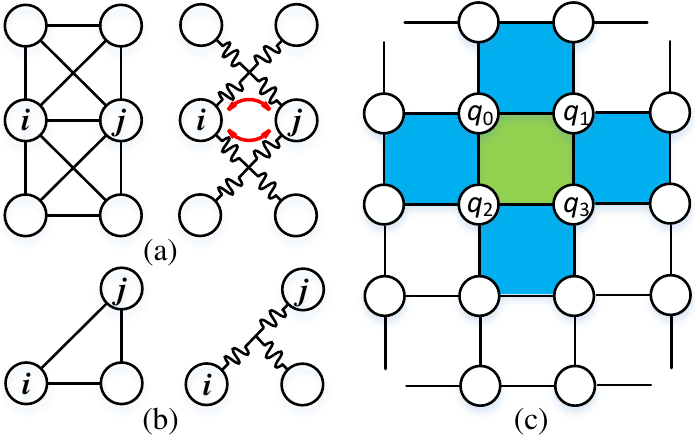}
\captionsetup{justification=centering}
\caption{(a) Prohibited Condition (b) Corner Case \\(c) Filtered Weight}
\label{fig:busselection}
\end{figure}

\begin{algorithm}[t]
\SetAlgoLined
 
\KwIn{Geometric coordinates of placed qubits, \csm, Maximum number of 4-qubit buses $K$}
\KwOut{Locations of 4-qubit Buses}
Calculate the cross coupling weight for each square\;
\While{$K>0$}{
   \tcp{Select one square in each iteration}
    \For{square(i, j) in all squares}{
       
        \textit{filtered\_weight\textit{(i, j)} $=$ weight\textit{(i, j)} - weight\textit{(i+1, j)} - weight\textit{(i, j+1)} - weight\textit{(i-1, j)} - weight\textit{(i, j-1)}}\;
    }
   
    \If{no square available for 4-qubit bus}{
        Break\;
    }
     Select the square with the highest $filtered\_weight$\;
     Set the weights of squares \textit{(i+1, j)}, \textit{(i, j+1)}, \textit{(i-1, j)}, and \textit{(i, j-1)} to be 0 and mark them to be blocked\;
     $K = K - 1$\; 
}

\caption{4-qubit Bus Selection}
\label{alg:connection}
\end{algorithm}

This paper simplifies the connection design problem by considering two types of common buses, 2-qubit bus and 4-qubit bus (shown in Figure~\ref{fig:scqubit}).
These two types of buses naturally fit in the 2D lattice qubit layout and can be easily fabricated because at most 4 nearby qubits are connected by one bus.
After placing the qubits on a 2D lattice in the first step, 2-qubit buses can be directly generated on the edges that connect two occupied nodes but the qubits on a diagonal of a 4-qubit square can never be connected with only 2-qubit buses.
Replacing some 2-qubit buses with 4-qubit buses could provide more qubit connection by trading in yield rate while it is not yet clear where to apply the 4-qubit buses can achieve the Pareto-optimal results.
The bus selection subroutine was proposed to identify the locations for 4-qubit buses.
Other potential bus designs are left as future research directions and will be discussed in Section~\ref{sec:futurework}.

Instead of considering the nodes in a 2D lattice, we consider the squares that are naturally formed by the edges in the 2D lattice.
Each square can be configured to 2-qubit bus or 4-qubit bus.
Now the problem is on which squares we should use 4-qubit buses.
The size of search space, even for this 4-qubit bus square selection problem, is still $O(exp(N))$.
But the simplification allows us to design high-quality heuristics to guide the selection.
Before introducing our solution, one additional prohibited condition must be considered. 

\textbf{Prohibited Condition} 
One physical constraint that we must consider when applying 4-qubit buses is that we cannot have 4-qubit buses in two adjacent squares.
The reason is explained with the example in Figure~\ref{fig:busselection} (a).
Suppose we have two adjacent squares and both of them are using 4-qubit buses.
Then there will be two physical connections between qubit $i$ and $j$.
When we use one of the connections, the other one will bring unexpected effects so that employing 4-qubit bus in one square will immediately block using 4-qubit buses in any of its adjacent squares.




Considering the physical constraints mentioned above, the objectives of this step are summarized as follows: 
\begin{itemize}
    \item Since adding more qubit connections will increase the probability of frequency collision and lower the yield, we hope to apply 4-qubit buses on those squares that can benefit the performance most.
    In other words, the additional connections are expected to directly support as many two-qubits gates as possible.
    \item Applying 4-qubit bus in one square will block adjacent squares, making it impossible to directly support some two-qubit gates in those blocked squares.
    This effect should also be considered when selecting the 4-qubit squares.
\end{itemize}


We propose a 4-qubit bus selection algorithm to select some squares for 4-qubit buses~(pseudocode shown in Algorithm~\ref{alg:connection}).
In each iteration, one square that could benefit most from a 4-qubit bus will be selected.
Users can specify the maximum number of 4-qubit buses they hope to have.
By varying the number of selected squares,  a series of architectures can be generated with a trade-off between yield and performance.

To find the most fitting square, we first need to calculate how much one square could benefit from a 4-qubit bus. 
Since the difference between a 2-qubit bus square and a 4-qubit bus square is whether the qubit pairs on the diagonals are connected, we define the cross-coupling weight for each square as the sum of the coupling strength of the qubit pairs on the diagonals.
For the example in Figure~\ref{fig:busselection} (c), the cross-coupling weight of the green square is the coupling strength of $(q_0,q_3)$ plus that of $(q_1,q_2)$.
A corner case in the coupling weight computation is the square with only 3 qubits (shown in Figure~\ref{fig:busselection} (b)).
In such squares, 4-qubit buses can naturally reduce to 3-qubit buses which support coupling between any two of the three connected qubits.
The weight of a 3-qubit square is only the weight of logical coupling between the two qubits on one diagonal since the other diagonal only has one qubit.
For example, the weight of the 3-qubit square in Figure~\ref{fig:busselection} (b) is the $(i,j)$ element in the \csm.
Except for this small modification, 3-qubit squares are treated equally as other 4-qubit squares in our bus selection step.
This cross coupling weight can estimate the potential benefit of applying 4-qubit bus in one square and realize the first objective.

However, the cross-coupling weight is not accurate enough to evaluate the benefit of 4-qubit for a square because the prohibited condition is not yet considered.
We design a filter to apply this constraint.
For each square, the filtered weight is its original cross-coupling weight minus all its neighbors' weights.
For example in Figure~\ref{fig:busselection} (c), the filtered weight of the green square is its original weight minus the weights of the four blue squares.
This filter can take the prohibited condition into consideration and achieve the second objective.

After applying the filter, we will select one square with the highest filtered weight.
Then we will label the selected square and its adjacent neighbors so that it will no longer be available for future 4-qubit buses.
We also change their weights to zero because they should not affect the 4-qubit selection among the remaining squares.
The algorithm will iterate again to select the next square until there are not more squares available or we have already applied enough number of 4-qubit buses.



\subsection{Frequency Allocation}

After the two steps above, we now have a complete coupling topology design of a superconducting quantum processor.
In the third step, we need to designate the pre-fabrication frequency of each qubit.
IBM's 5-frequency scheme is a regular frequency designation~\cite{rosenblatt2019enablement}. 
However, the generated qubit layout and connection in our design flow can be irregular since more hardware sources are invested in locations that can benefit the performance most.
Thus, we need a more flexible frequency allocation scheme to leverage this unbalanced qubit layout and connection.
The objective of this step is to minimize the probability of post-fabrication frequency collision and improve the yield rate. 
The physical constraints are the frequency collision conditions in Figure~\ref{fig:collisions}.

\begin{figure}[b]
\centering
\includegraphics[width=1.0\columnwidth]{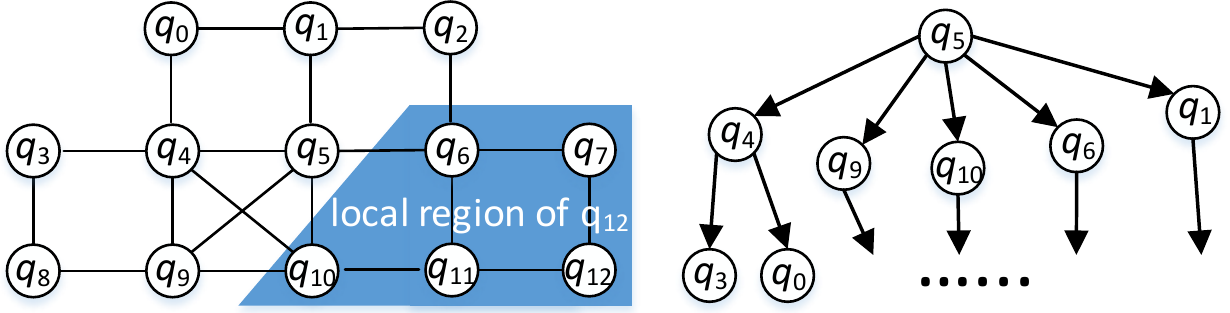}
\vspace{-15pt}
\caption{Breath First Frequency Allocation}
\label{fig:freqalloc}
\end{figure}

Finding the qubit frequency allocation plan to maximize the yield rate is a hard problem.
The complex collision conditions make it difficult to find an analytic expression for the yield rate and a brute-force search over all possible frequency configurations will be very time-consuming.
For example, if there are $M$ candidate frequencies for each qubit and we have $N$ qubits in total, the total number of possible frequency configurations is $M^N$. 
For each of these potential configurations, we need to run a yield simulation (introduced in Section~\ref{sec:yieldsim}) and then select the one with maximal yield rate.
This method is not acceptable due to its high complexity.
We propose to optimize the qubit frequency allocation algorithm based on the facts that 1) the physical qubits in the geometric center of the qubit lattice are more likely to involve in a frequency collision since they usually have more qubit connections, and 2) frequency collision only happens among nearby qubits.

\begin{algorithm}[t]
\SetAlgoLined
 
\KwIn{Qubit Location and Connection}
\KwOut{Frequency Configuration of Each Qubit}
\SetKwRepeat{Do}{do}{while}
Select the qubit in the geometric center of the placed qubits and set its frequency to be the middle of the allowed frequency range\;

\Repeat{the frequencies of all qubits are determined}{
    Find the next qubit $q_i$ in breadth-first traversal order\;
    \For{$temp\_freq$ in all frequency samples}{
        Set the frequency of $q_i$ to be $temp\_freq$\;
        Simulate the yield rate within $q_i$'s local region\;
    }
    Assign the frequency with maximal yield rate to $q_i$\;
}

\caption{Frequency Allocation}
\label{alg:freq}
\end{algorithm}

Our algorithm determines the qubit frequencies from the center to the periphery (pseudocode shown in Algorithm~\ref{alg:freq}).
Since this step is purely about hardware, the input of our algorithm is only the qubit location and connection generated from the previous two subroutines.
To reduce the manufacturing difficulty and help prevent the collision condition 4, we follow the convention from IBM and set an allowed frequency interval $5.00GHz$ to $5.34GHz$.
All pre-fabrication frequencies are limited within this interval.
First, we locate the qubit that is closest to the center of the qubit lattice and assign its frequency to be the center of the allowed frequency interval.
Then we apply breadth-first traversal on the coupling graph from the first qubit in the center.
For example, $q_5$ is the center qubit in the example shown in Figure~\ref{fig:freqalloc}.
In the breadth-first traversal, we will first access $q_{4,9,10,6,1}$ as shown on the right.
Each time we access one new qubit, we will immediately determine its frequency.
A list of candidate frequencies is prepared.
In this paper, the candidate frequencies are $5.00, 5.01, 5.02, \dots, 5.33, 5.34GHz$ to achieve an accuracy of $0.01GHz$.
We can also have more candidate frequencies but it will take more time to evaluate all of them.

To evaluate a candidate frequency on a new qubit, we temporarily assign the candidate frequency to the new qubit and then simulate the yield rate within its local region.
The local region of a qubit is defined as a sub-graph of the original chip coupling graph in which a qubit may collide with the new qubit.
For example in Figure~\ref{fig:freqalloc}, when we are searching for the best frequency of $q_{12}$, the local region is marked in blue. Qubits not in this region like $q_5$ cannot collide with $q_{12}$.
We will select the frequency with the maximal yield rate and assign it to the new qubit.
Now the time complexity of the frequency allocation algorithm is $O(MN)$ where $M$ is the number of candidate frequencies and $N$ is the number of qubits.

\subsubsection{Yield Simulation} \label{sec:yieldsim}
We developed a yield simulator based on IBM's yield model~\cite{rosenblatt2019enablement, brink2018device}.
The fabrication process can be modeled by adding a Gaussian noise $N(0,\sigma)$ to the pre-fabrication frequency of a qubit to generate its post-fabrication frequency where $\sigma$ is the fabrication precision parameter.
For a given superconducting quantum processor design, we estimate its yield rate through Monte Carlo simulation.
Each time we will simulate if one fabrication is successful. 
We first generate the post-fabrication frequencies by adding a random noise sampled from Gaussian distribution mentioned above.
Then we check if any frequency collision condition listed in Figure~\ref{fig:collisions} occurs in the post-fabrication frequencies.
If so, this fabrication fails. Otherwise, it is successful.
All possible cases are taken into account.
For example, we will examine the two frequencies of all connected physical qubit pairs for condition 1, 2, 3, and 4.
If they meet any one of the inequalities of the conditions, frequency collision is considered to occur in this simulation.
This simulation process is repeated many times.
The yield rate can be estimated by the ratio between the number of successful simulations and the total number of simulations.


\section{Evaluation}\label{sec:evaluation}


To demonstrate that the proposed application-specific architecture design flow can deliver hardware designs with better Pareto-optimal results in terms of performance and yield rate, we conduct experiments over various benchmarks to show not only the overall improvement  
but also the breakdown of benefits from each of our hardware design subroutines.


\subsection{Experiment Setup}


\textbf{Benchmarks}
Twelve quantum programs are collected from IBM's QISKit~\cite{IBMqiskit} and RevLib~\cite{wille2008revlib}, or compiled from ScaffCC~\cite{javadiabhari2015scaffcc}. 
These benchmarks cover several important domains (e.g., simulation, arithmetic) and have various sizes (from 7- to 16-qubit) for a versatility test of the proposed design flow. 

\textbf{Metrics}
To evaluate the efficiency of an architecture, we need both the yield rate and performance.
An architecture with a higher yield rate can be successfully fabricated with fewer attempts, indicating a lower hardware cost.
In our experiments, the yield rate is simulated with IBM's yield model~\cite{rosenblatt2019enablement, brink2018device} as introduced in Section~\ref{sec:yieldsim}.
For the performance evaluation, we adopt the total post-mapping gate count metric widely used in previous studies~\cite{zulehner2018efficient, siraichi2018qubit, li2019tackling}.
More gates lead to longer execution time and a larger probability of error on QC devices. 
If a hardware architecture could execute the program with fewer gates, then its performance is considered to be better.

\textbf{Yield Simulation Configuration}
The number of trials in the Monte-Carlo simulation for each architecture is 10,000, which is $10\times$ of that used in IBM's experiments~\cite{brink2018device, hutchings2017tunable, chamberland2019topological} to ensure the simulation accuracy.
The fabrication precision parameter $\sigma$ is set to be $30MHz$, a realistic extrapolation of progress in hardware by IBM~\cite{rosenblatt2019enablement, chamberland2019topological}. 
IBM has improved the $\sigma$ from $200MHz$~\cite{rosenblatt2017variability} to $130MHz$~\cite{rosenblatt2019enablement} in the last few years and $30MHz$ is a reasonable projection to achieve a useful yield as predicted by IBM~\cite{chamberland2019topological}.



\subsection{Experiment Methodology}
To illustrate the benefit of our design flow, five experiment configurations are designed to show the overall improvement and the performance/yield trade-off gain at each of the three subroutines in Section~\ref{sec:organization}.
Among them, \textbf{\textsf{ibm}} is a set of general-purpose architectures from IBM and they are not tailored for any applications.
The remaining four configurations are application-specific architectures generated by the entire or part of the proposed design flow.

\begin{figure}[t]
\centering
\includegraphics[width=1.0\columnwidth]{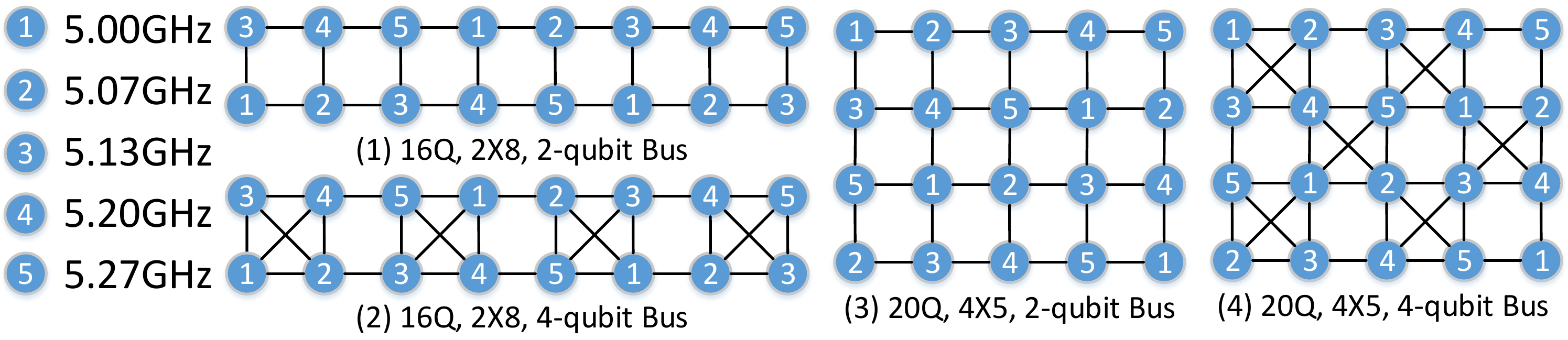}
\vspace{-15pt}
\captionsetup{justification=centering}
\caption{Baseline Qubit Frequency, Layout, \\ and Connection Designs}
\label{fig:baseline}
\end{figure}

\textbf{\textsf{ibm}} We use IBM's design scheme as the baseline configuration.
It has two layout options, a 2$\times$8 lattice with 16 qubits, and a 4$\times$5 lattice with 20 qubits.
The qubit connection design can be either 2-qubit bus only or using 4-qubit buses as many as possible. 
In total, there are four architectures combining the layout and connection options~(shown in Figure~\ref{fig:baseline}).
The frequency allocation scheme is a 5-frequency scheme~\cite{rosenblatt2019enablement, chamberland2019topological}. 
The five frequencies are an arithmetic progression from $5GHz$ to $5.27GHz$ and their arrangement is also in Figure~\ref{fig:baseline}. 


\textbf{\textsf{eff-full}} We apply all three subroutines and generate a series of efficient superconducting quantum processor architectures by varying the number of 4-qubit buses. 
The number of designs we can obtain for a quantum program depends on the number of qubits as more qubits can provide more squares to apply 4-qubit buses in the generated layout.
This experiment can show the overall architecture design improvement when comparing with the baseline \textbf{\textsf{ibm}}.

\textbf{\textsf{eff-5-freq}} We only apply the first two subroutines to generate qubit layout and connection design but the frequency allocation is done with IBM's 5-frequency scheme. The yield benefit from the proposed frequency allocation algorithm can be demonstrated by comparing with results from \textbf{\textsf{eff-full}}.  

\textbf{\textsf{eff-rd-bus}} We keep the first and the third subroutines but randomly select some squares to employ 4-qubit buses with the prohibited condition constraint satisfied. 
This will demonstrate the effect of our filtered-weight-based 4-qubit bus selection algorithm by comparing with results from \textbf{\textsf{eff-full}}.

\textbf{\textsf{eff-layout-only}} We apply our profiling method and perform a layout design. 
The connection design has two options. 
One is only using 2-qubit buses.
The other is using 4-qubit buses as much as possible.
The frequency design follows the baseline \textbf{\textsf{ibm}}.
The benefit of our layout optimization can be shown when comparing with the results from \textbf{\textsf{ibm}}.

For each benchmark, we run all the five configurations to generate different superconducting quantum processor architectures with different yield rates. 
Then we apply one state-of-the-art qubit mapping algorithm~\cite{li2019tackling} on these architectures to obtain the total number of gates when running the generated or baseline architectures.




\subsection{Overall Improvement}
\begin{figure*}[t]
\centering
\includegraphics[width=2.0\columnwidth,height=1.35\columnwidth]{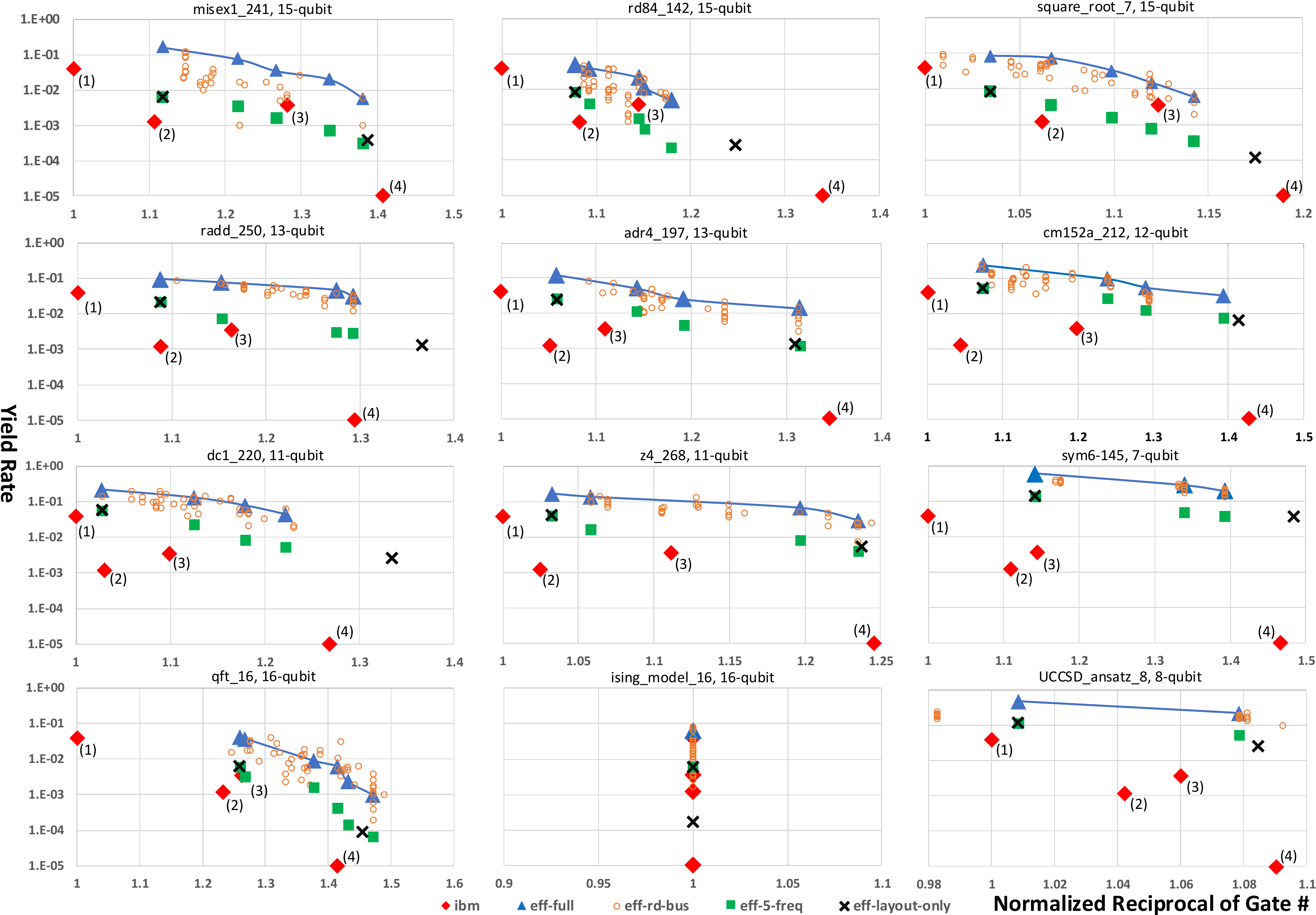}
\vspace{-5pt}
\captionsetup{justification=centering}
\caption{Yield v.s. Normalized Reciprocal of Post-mapping Gate Count}
\label{fig:result}
\end{figure*}

Figure~\ref{fig:result} shows the result of yield and performance for all benchmarks and the five experiment configurations.
There are 12 subfigures and one subfigure contains the results of the five experiment configurations for one benchmark.
The X-axis represents the normalized reciprocal of post-mapping gate count and data points on the \textbf{right} have better performance. 
The Y-axis represents the yield rate and data points on the \textbf{top} have higher yield rates.
The legend at the bottom of Figure~\ref{fig:result} shows the markers for the five configurations.
The data points for the four designs in the baseline are labeled by (1), (2), (3), and (4), according to Figure~\ref{fig:baseline}.

\textbf{Optimality} A series of architectures with  better Pareto-optimal results of performance and yield since the data of \textbf{\textsf{eff-full}} is on the upper right of \textbf{\textsf{ibm}}.
The most simplified designs (the most left top blue data point) generated by our design flow outperforms the 16-qubit baseline design without 4-qubit buses in both performance ($\sim 7.7\%$) and yield rate ($\sim 4\times$).
Compared with the 16-qubit baseline with four 4-qubit buses, we achieve over $100\times$ better yield rate with $<1\%$ performance loss.
On the other side, compared with IBM's 20-qubit chip design with six 4-qubit buses (the baseline design with the most hardware resources), the designs with the maximum number of 4-qubit buses generated from our design flow have over $1000\times$ yield rate improvement on average with only about $3.5\%$ performance loss. 

\textbf{Controllability} 
The proposed design flow can easily control the trade-off between yield and performance by only changing the number of 4-qubit buses without traversing across, or sampling a large number of designs in, the entire search space.
Depending on the number of qubits in different target programs, we can trade in around $10\times\sim50\times$ yield rate for $10\%\sim33\%$ performance improvement.


\subsubsection{Special Case}
The results of \textit{ising\_model}  are significantly different because the logical qubit coupling in this benchmark forms a chain structure.
The mapping algorithm can always find the perfect initial mapping without inserting additional operations.
As a result, the post-mapping gate count is the same for all tested hardware architectures.
All data points for this program lie in one vertical line.
Only one architecture is generated from our design flow because there is no need to add 4-qubit bus.
All the two-qubit gates can be executed through the edges on the 2D lattice. There are no two-qubit gates applied on two qubits on a diagonal because of the chain coupling structure.
In this case, 4-qubit buses can only lower the yield rate without improving the performance.

\subsection{Effects from Individual Subroutines}
The overall improvement has already been discussed, but one interesting question is how much improvement the layout and connection optimization contribute and how much comes from the optimized yield allocation directly.
The five configurations decouple the proposed design flow and provide a breakdown of the effect of individual subroutines.

\subsubsection{Effect of Layout Design}
The difference between \textbf{\textsf{ibm}} and \textbf{\textsf{eff-layout-only}} illustrates the effect of layout design since the rest two subroutines are  the same.
An architecture with more hardware resources is expected to provide higher performance by allowing more flexibility in qubit mapping.
But our optimized layout design could use comparable or fewer hardware resources while the performance can be even better. 
For example, we compare the 2-qubit bus only data point (the upper left one) with the 16-qubit baseline with four 4-qubit buses (labeled by (2) in each subfigure).
\textbf{\textsf{eff-layout-only}} provides better or comparable performance most of the time with about $35\times$ yield improvement on average.
The improvement at this step depends on the program size and programs with fewer qubits will use fewer qubits and connections in an optimized architecture.
This result proves that our layout design could generate qubit layout with high performance but using much fewer hardware resource for different programs.

\subsubsection{4-qubit Bus Selection Quality} 
By comparing the results from \textbf{\textsf{eff-full}} and \textbf{\textsf{eff-rd-bus}}, we can see that the architectures generated from our bus selection algorithm are better than that of random selection in trading in yield for performance most of the time.
The data points of \textbf{\textsf{eff-rd-bus}} reveal the distribution of the yield and performance sampled from random bus designs.
Note that the performance of \textbf{\textsf{eff-rd-bus}} is usually confined by the two data points in \textbf{\textsf{eff-layout-only}} because adding connections can improve the performance most of the time.
For most benchmarks except \textit{qft}, the results from \textbf{\textsf{eff-full}} are close to the upper bound formulated by the random samples, which shows that our weight-based bus selection could generate a series of near Pareto-optimal hardware architectures with various numbers of qubit connections. 

The result of \textit{qft} is much worse than that of other programs due to the unique uniform two-qubit gate pattern in this program.
The number of two-qubit gates between arbitrary two logical qubits is always two in \textit{qft}, which makes all the logical qubit pairs are the same in the sense the coupling strength during profiling.
Then in bus selection subroutine, all the squares share the same weight and the weight-based selection is the same as random selection.


For the two small benchmarks, \textit{sym6} and \textit{UCCSD\_ansatz}, the number of available squares in the generated qubit layout is small and there are very few options when applying 4-qubit buses.
Therefore, most of the architectures generated from the random 4-qubit bus selection are the same as those from the proposed design flow, which makes the results from \textbf{\textsf{eff-full}} and \textbf{\textsf{eff-rd-bus}} very close.


\subsubsection{Frequency Allocation Optimization}
By comparing \textbf{\textsf{eff-full}} and \textbf{\textsf{eff-5-freq}}, we can see that the proposed frequency allocation algorithm provides about 10$\times$ yield rate improvement on average.
This improvement is slightly worse when the yield from the baseline 5-frequency is already high, e.g., results from \textit{sym6} and \textit{UCCSD\_ansatz}.
The fabrication variance makes the ideal yield 100\% unreachable and it is hard to optimize yield when it is already high.


\section{Discussion}\label{sec:futurework}
This paper studies application-specific efficient superconducting quantum processor design.  
In particular, we formalize the architecture design for superconducting quantum processors with three key steps, each of which comes with an optimization subroutine.
This is the first attempt, to the best of our knowledge, to identify the optimization opportunity from the architecture level to push forward the balance between QC performance and hardware yield rate. 
Effort towards this direction can be of significant demand in the near term QC with limited computation resource and immature fabrication technology. 

Although we show that improved Pareto-optimal designs can be generated with a static program analysis and three optimized design algorithms, several future research directions can be explored as with any initial research.

\textbf{Improving Profiling Method}
This paper focused on the logical qubit coupling topology in a quantum program but other patterns may also be leveraged.
We omitted the temporal information of the two-qubit gates and all information about other program components.
But the locations of two-qubit gates in a quantum program may also be leveraged for finer-grained evaluation of the coupling strength for different logical qubit pairs at different times during the execution.
The single-qubit patterns can also help with the basic gate set design.


\textbf{Exploring More Design Space}
In the proposed design flow, the number of physical qubits is  the same as that of logical qubits for higher yield rate.
However, we can still add auxiliary physical qubits since they can also be used during the qubit routing, trading in more yield rate for higher performance.
How to add auxiliary qubit to appropriate locations and how to connect them are interesting problems to explore in the future.
To ensure modularity and scalability, the qubits are forced to be embedded in a 2D lattice and only consider two types of buses lying in the lattice.
However, the qubit placement and connection could be more flexible if we trade in part of the scalability.
For example, one bus could also connect more than four qubits~\cite{ghosh2013high}.
The design space in this direction is not yet explored.

\textbf{Optimizing Frequency Allocation}
This paper tried to optimize the qubit frequency selection from the center to periphery and only searched for the optimal frequency for one qubit, resulting in a sub-optimal frequency allocation.
A global optimization like formal methods can be explored to further optimize the frequency allocation result.
One alternative approach to resolve the frequency collision issue is to use flux-tunable transmon qubits~\cite{kelly2015state}, of which the frequencies can be dynamically tuned with additional control signals.
The design trade-off of different types of qubits is not yet explored and additional signals bring more noise and increase the control complexity.
The proposed design flow is still valuable even with frequency-tunable qubits because the simplified architectures with fewer the on-chip connections can not only reduce the fabrication complexity but also benefit the overall performance by lowering the crosstalk error.

\section{Related Work}\label{sec:relatedwork}

This paper ranges across multiple topics, i.e., program profiling, superconducting processor design, application-specific design, qubit mapping.
We briefly introduce related work for all of them.

\textbf{Application-specific Design}
The closest related work is SPARQS, a superconducting planar architecture proposed by Wilhelm \textit{et al.}~\cite{dallaire2016quantum,liebermann2017implementation} targeting a specific Fermi-Hubbard model simulation program.
However, they only provide an implementation-independent design from theoretical physics level.
This paper formalizes a systematic end-to-end design flow with automatic program profiling and realistic physical constraints included, for the first time.
With no limitation on the target program, we can generate a series of Pareto-optimal hardware architecture designs in a controllable way.

\textbf{Quantum Program Profiling and Analysis}
Program profiling and analysis are very important for software and compiler optimization.
Previous works on quantum program analysis~\cite{javadiabhari2015scaffcc,ying2010quantum,ying2013verification,ying2013reachability,honda2015analysis,perdrix2008quantum} have studied entanglement, termination, non-cloning checking, etc. 
The profiling method in this paper is proposed to guide the hardware design, fulfilling a different goal.

\textbf{Superconducting Quantum Processors}
As one of the most promising candidate technology to implement QC, superconducting quantum techniques have been employed in two mainstream QC computation models.
The circuit model based processors~\cite{IBMdevice,RigettiQPU,Google72Q} support quantum circuit model \cite{nielsen2010quantum} and the quantum annealers~\cite{DWaveDoc} can implement adiabatic QC~\cite{farhi2000quantum}.
Their programming model and hardware architecture are different for these two QC approaches.
The design flow in this paper is proposed for circuit model based quantum processors while efficient quantum annealer design can be a future research direction. 

\textbf{Qubit Mapping}
Formal and heuristic methods have been attempted to solve this problem~\cite{venturelli2018compiling,shafaei2014qubit,li2019tackling,zulehner2018efficient,siraichi2018qubit} and minimize the total gate count.
Recently several studies~\cite{tannu2019not,ash2019qure,murali2019noise} have applied the actual gate error rates for fine-grained optimization.
All these optimizations are pure software-level modification.
This paper attempts to improve the performance by reducing the mapping overhead from the hardware level.
We adopt the gate count metric to estimate the mapping overhead since our experiments are performed on artificial hardware architectures. 


\section{Conclusion}\label{sec:conclusion}

The demand for larger computation capability in a superconducting quantum processor naturally calls for more hardware resources which will also increase the design complexity and lower the yield rate.
This paper explored application-specific architecture design for superconducting quantum processors to achieve both high performance and higher yield rate.
Gate patterns in a quantum program can be extracted by the proposed profiling method and then utilized in the follow-up hardware architecture design.
Three subroutines are designed to generate the qubit layout, connection, and frequency respectively with physical constraints taken into consideration.
Experimental results show that the proposed design flow could deliver architectures with both high yield rate and performance automatically for different applications except those with extremely special gate patterns.


\bibliographystyle{unsrt}
\bibliography{main}

\end{document}